\documentclass[prb,showpacs,floatfix,twocolumn]{revtex4}
\usepackage{amsfonts}
\usepackage{stmaryrd}
\usepackage{bbm}
\usepackage{mathrsfs}
\usepackage{tipa}
\usepackage{amssymb}
\usepackage{txfonts}
\usepackage{graphicx}
\usepackage{dcolumn}
\begin{document}

\newcommand*{\cm}{cm$^{-1}$\,}


\title{Effect of disorder in the charge-density-wave compounds LaTe$_{1.95}$
and CeTe$_{1.95-x}$Se$_x$ (x=0 and 0.16) as revealed by optical spectroscopy}


\author{Y. Huang}
\author{B. F. Hu}
\author{T. Dong}
\author{A. F. Fang}
\author{P. Zheng}
\author{N. L. Wang}

\affiliation{Beijing National Laboratory for Condensed Matter
Physics, Institute of Physics, Chinese Academy of Sciences,
Beijing 100190, People's Republic of China}
%


\begin{abstract}
We present optical spectroscopy measurements on rare-earth ditelluride single crystals of
LaTe$_{1.95}$ and CeTe$_{1.95-x}$Se$_x$ (x=0 and 0.16). The measurements reveal
formation of charge density wave energy gaps
at rather high energy levels, \emph{e.g.} 2$\Delta\sim$ 8500 \cm for
LaTe$_{1.95}$, and 6800 \cm for CeTe$_{1.95}$. More strikingly, the study reveals that, different
from the rare-earth tri-tellurides, the Te vacancies and disorder effect
play a key role in the low-energy charge excitations of ditelluride systems.
Although an eminent peak is observed between 800 and 1500 \cm in conductivity spectra for
LaTe$_{1.95}$, and CeTe$_{1.95-x}$Se$_x$ (x=0. 0.16), our analysis indicates that
it could not be attributed to the
formation of a small energy gap, instead it could be well accounted for
by the localization modified Drude model. Our study also indicates that the low-tempreature
optical spectroscopic features are distinctly different
from a semiconducting CDW state with entirely gapped Fermi surfaces.

\end{abstract}


\maketitle

\section{Introduction}

Charge density wave (CDW) is a collective quantum phenomenon in solids and a
subject of considerable interest in condensed matter physics. Most CDW states are
driven by the nesting topology of Fermi surface (FS),
i.e. the matching of sections of FS to
others by a wave vector 2\textbf{k}$_F$, where the electronic susceptibility has a divergence.
A single particle energy gap opens in
the nested regions of the Fermi surfaces at the transition,
which leads to the lowering of the electronic energies of the
system. Coupling to the lattice, the development of CDW state would also
cause a lattice distortion with the modulation wave vector of superstructure
matching with the FS nesting wave vector.\cite{G.Gruner}

The nesting condition is easily realized in low-dimensional electronic
systems. In one-dimensional (1D) CDW systems, a perfect nesting can be
realized and the FS could be fully gapped. Then the systems become
insulating or semiconducting in the CDW phase. For 2D or 3D CDW systems,
a perfect nesting of the entire FSs could hardly be fulfilled. In this circumstance,
the CDW energy gap forms only on the partially nested region of FSs. Due to the presence of
ungapped region of FSs, the system would remain metallic even in the CDW state.

The rare-earth polychalcogenides RTe$_n$ (where R is La or rare earth element,
n=2, 2.5, 3) are prototype CDW-bearing materials. These systems have layered
structures, consisting of corrugated rare-earth-chalcogen slabs alternated with planar
chalcogen Te square lattice. R in the compound is
trivalent, donating three electrons to the system. They completely
fill the Te p orbitals in the RTe slabs, but partially those Te p
orbitals in the square Te-layers
\cite{structure2,ARPESCeTe3400meV}. Metallic conduction occurs in
the Te layers, leading to highly anisotropic transport properties
\cite{structure2,material,TEMRTe3}. Nested regions of FSs were indicated by both band structure calculations
and ARPES measurements \cite{Kikuchi,B.I.Minsuper,J.Lcaculate,fisherRTe2},
which has been well characterized as the origin of CDW formation.
Pressure-induced superconductivity was also found in several systems of both rare-earth tri- and di-tellurides
in the family \cite{DiMasi,superconduct}, yielding good candidates for investigating the competition between superconductivity
and CDW orders. Among the family, the rare-earth tri-telluride RTe$_3$,
which consists of double Te layers between insulating
corrugated RTe slabs, were widely studied. Two energy gaps with different energy scales were observed \cite{optCeTe3,optErTe3},
which were considered as driven by two different nesting wave vectors present in the FS topology.
Similar to other 2D CDW systems, the ungapped regions of FSs are always present in RTe$_3$ and
the materials remain metallic in CDW state. Compared with rare-earth
tri-telluride RTe$_3$, much less work has been
done on the rare-earth ditelluride RTe$_2$
which consists of single Te layers between insulating corrugated RTe slab (see inset
of Fig. 1). The reported CDW gap structures by ARPES measurement are rather controversial.
Shin \emph{et al.} performed ARPES measurements on LaTe$_{1.95}$ and CeTe$_2$ and found that for both compounds
the inner FSs center at $\Gamma$ point are almost fully gapped with E$_g$=600 meV while the outer FSs
are only partially gapped with E$_g$=100 meV.\cite{fisherRTe2} On the other hand, Garcia \emph{et al.} investigated LaTe$_2$
compound and found that the entire inner and outer FSs are gapped by a surprisingly small
energy scale of E$_g$=50 meV as determined from the leading edge shift \cite{ARPESLaTe2}. They claimed that CDW gap size
decreases dramatically as the number of the Te layers reduces from two (RTe$_3$) to one (RTe$_2$)
and the LaTe$_2$ would be the first proven instance of semiconducting quasi-2D CDW material \cite{ARPESLaTe2}.

It would be essential to clarify the issue by performing different spectroscopic
measurements. It should be noted that, unlike the case of rare-earth tri-tellurides
where the conducting Te layers are free from defects, Te vacancies in Te layers were found
to be present in most reported work on rare-earth ditelluride RTe$_2$ compounds.
Special care has to be taken on the sample characterizations.
Optical spectroscopy is a powerful bulk sensitive technique to
detect the energy gaps in ordered state and yields a great wealth of
information in CDW systems. Here we present optical spectroscopic
measurements on LaTe$_{1.95}$, pure and Se-doped CeTe$_{1.95}$
single crystals. Our measurement indicates clearly the formation of CDW gap
structure at rather high energy level with 2$\Delta\sim$ 8500 \cm ($\sim$1.06 eV) for
LaTe$_{1.95}$. The energy scale of the CDW gap is gradually reduced
for the pure and Se-doped CeTe$_{1.95}$ samples.
Although a pronounced peak at low energy scale, between 800$\sim$1500 \cm (0.1$\sim$0.2 eV)
for LaTe$_{1.95}$ and CeTe$_{1.95-x}$Se$_x$ (x=0, 0.16), is also observed,
our study suggests that the low energy excitations are dominantly contributed
by the disorder effects due to the presence of Te vacancies in Te layers. The
experimental results are very different from the defect-free rare-earth tritelluride RTe$_3$
compounds where small CDW energy
gaps could be clearly indicated. Furthermore, the
spectral features are distinctly different from a semiconducting CDW state
with fully gapped Fermi surfaces.

\section{\label{sec:level2}EXPERIMENT AND RESULTS}

Single crystals of RTe$_{2-x}$ (R=La, Ce) have been grown by a self-flux
technique.\cite{fisherRTe2} The mixtures of rare-earth powders and
Te pieces in an atomic ratio from 0.16:0.84 to 0.18:0.82 were placed
in an alumina crucible and sealed in an evacuated quartz tube. The
mixture was heated up to $1150^{\circ}C$ and kept for one
day, then cooled down slowly to $1000^{\circ}C$ over a
period of 5 days. At the final temperature, the rest flux Te was
separated from single crystals in a centrifuge. Shiny and dark colored crystals were obtained.
The crystals were found to be air- and
moisture-sensitive. We also grew Se-doped single
crystals of CeTe$_2$ from similar process by changing the starting compositions
to Ce$_{0.18}$Te$_{0.76}$Se$_{0.06}$.

Figure 1 displays the X-ray diffraction pattern of CeTe$_{2-x}$ single crystals
at room temperature. The (0 0 $l$) diffraction peaks indicate a good c-axis characteristic.
The obtained c-axis lattice parameter is
c=9.11, which agrees well with the previous result.\cite{xrd}
The energy dispersive X-ray spectroscopy (EDX) analysis equipped with
the scanning electron microscope (SEM)
indicates that both the La- and Ce-based compounds have average compositions of
R:Te$\approx$1:1.95 (R=La, Ce). The Se-doped crystal has an average composition of
CeTe$_{1.79}$Se$_{0.16}$. Obviously, Te deficiencies are present
in the crystals although the self-flux method was reported to be effective on
reducing the Te vacancies.\cite{fisherRTe2} As we shall elaborate in this work, those Te vacancies
greatly affect the optical properties of RTe$_2$ systems.

\begin{figure}[t]
\centering
\includegraphics[clip,width=3.3in]{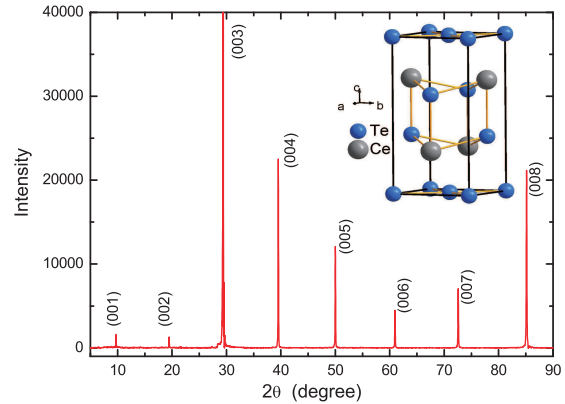}
\caption{(Color online) The (00l) x-ray diffraction pattern of
single-crystal CeTe$_{1.95}$ The strongest peak is only partially
displayed in order to show other s clearly.The inset shows the
crystal structure.}
\end{figure}

The temperature dependence of the in-plane (ac-plane) dc conductivity $\rho(T)$
was measured by a standard four-probe method in a quantum design
physical properties measurement system (PPMS) and plotted in Fig. 2.
Platinum wires were fixed on the sample using highly conducting silver
adhesive in the glove-box to avoid deterioration. The resistivity of
both CeTe$_{1.95}$ and LaTe$_{1.95}$ shows metallic behavior with relatively
high absolute values at base temperature in comparison with RTe$_3$. In an earlier
report by by Shin et al. \cite{fisherRTe2}, the resistivity of LaTe$_{1.95}$
shows an upturn at low temperature. The different behaviors could be attributed
to slightly different sample quality. Due to Te deficiencies, $\rho(T)$ values change
between crystals \cite{Resist,fisherRTe2}. In addition,
CeTe$_{1.95}$ compound shows a sharp feature at about 5K, which is
related to the anti-ferromagnetic ordering of spins from the
localized 4f electrons of Ce. On the other hand, the
resistivity behavior is rather different when Te was partially
substituted by Se. The resistivity of CeTe$_{1.95-x}$Se$_x$ (x=0.16) increases modestly
with decreasing temperature. An anomaly is surprisingly seen near 345 K both in
cooling and warming processes. We performed transmission electron microscopy (TEM)
measurement on this sample and found that the superlattice diffraction spots
disappear at temperatures above 345 K. Combined with the results from
optical study below, we believe that it could be ascribed to CDW phase
transition.

\begin{figure}[b]
\centering
\includegraphics[clip,width=3.3in]{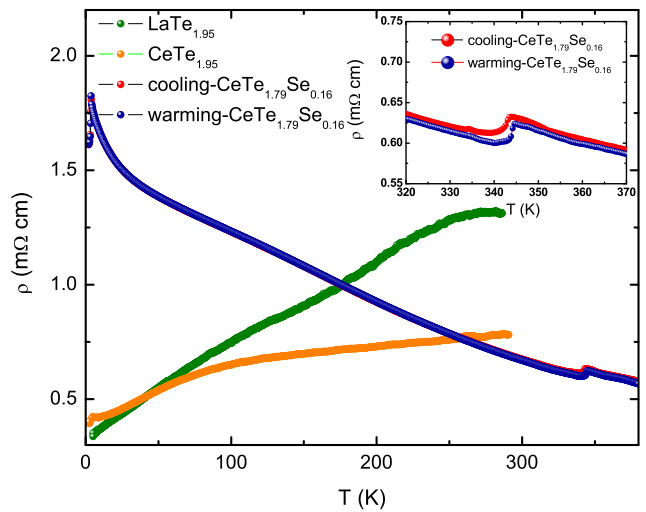}
\caption{(Color online) The temperature dependent in-plane
(ac-plane)resistivity, showing LaTe$_{1.95}$ ,CeTe$_{1.95}$ and
CeTe$_{1.79}$Se$_{0.16}$. Inset: the details of the anomaly of
CeTe$_{1.79}$Se$_{0.16}$ between 340k and 350k.}
\end{figure}

The optical reflectivity measurement was carried out on Bruker IFS
113v and 80v/s spectrometers in a frequency range from 40 to 25000
cm$^{-1}$. An \textit{in situ} gold and aluminium overcoating
technique was used to get the reflectivity R($\omega$).
Kramers-Kronig transformation of R($\omega$) is employed to get the
real part of the conductivity spectra $\sigma_1(\omega)$. A
Hagen-Rubens relation was used for the low frequency extrapolation.
A constant value of high frequency extrapolation was used up to
100000 \cm, above which an $\omega^{-4}$ relation was employed.

\begin{figure*}[t]
\centering
\includegraphics[clip,width=2.4in]{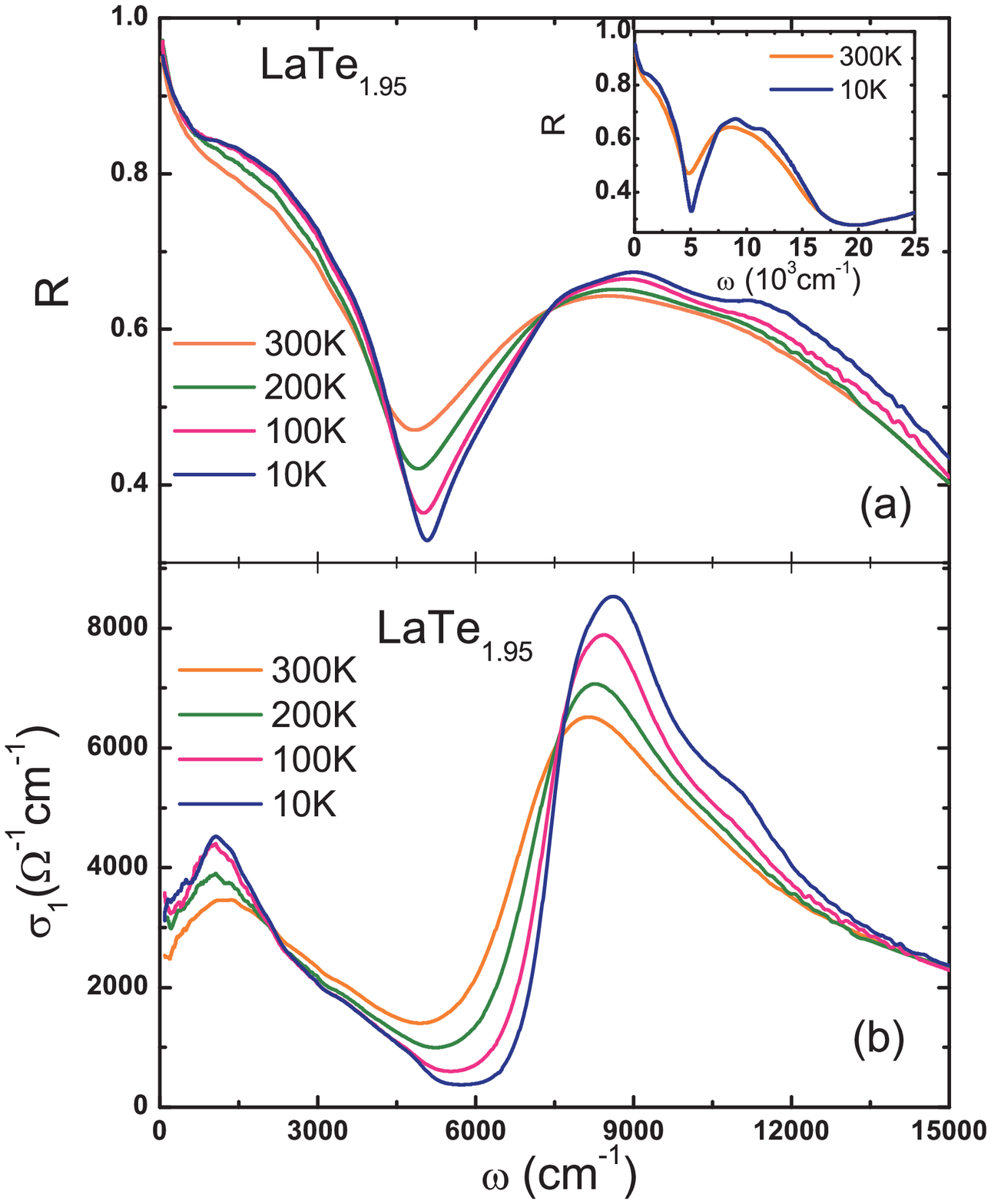}~
\includegraphics[clip,width=2.4in]{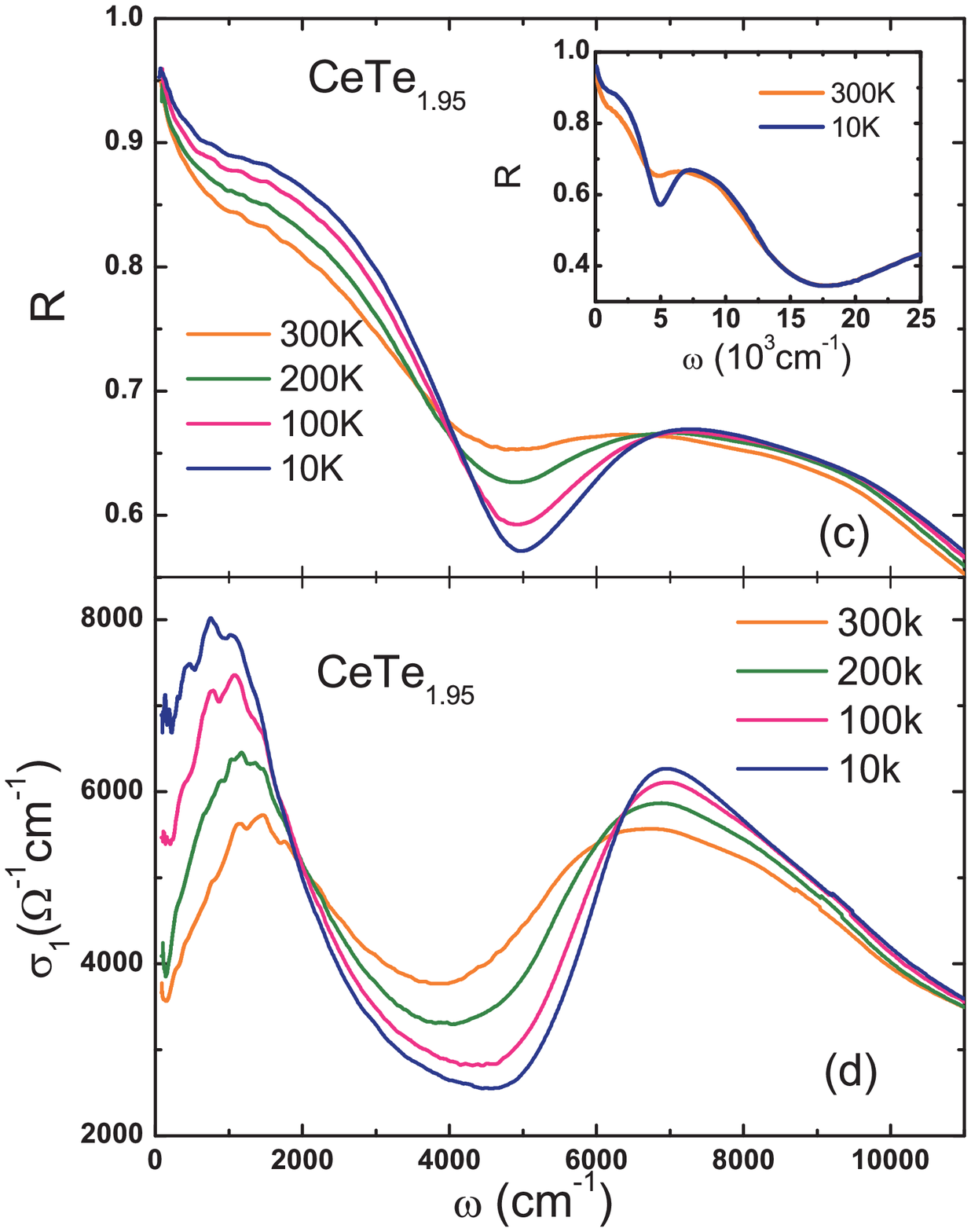}~
\includegraphics[clip,width=2.4in]{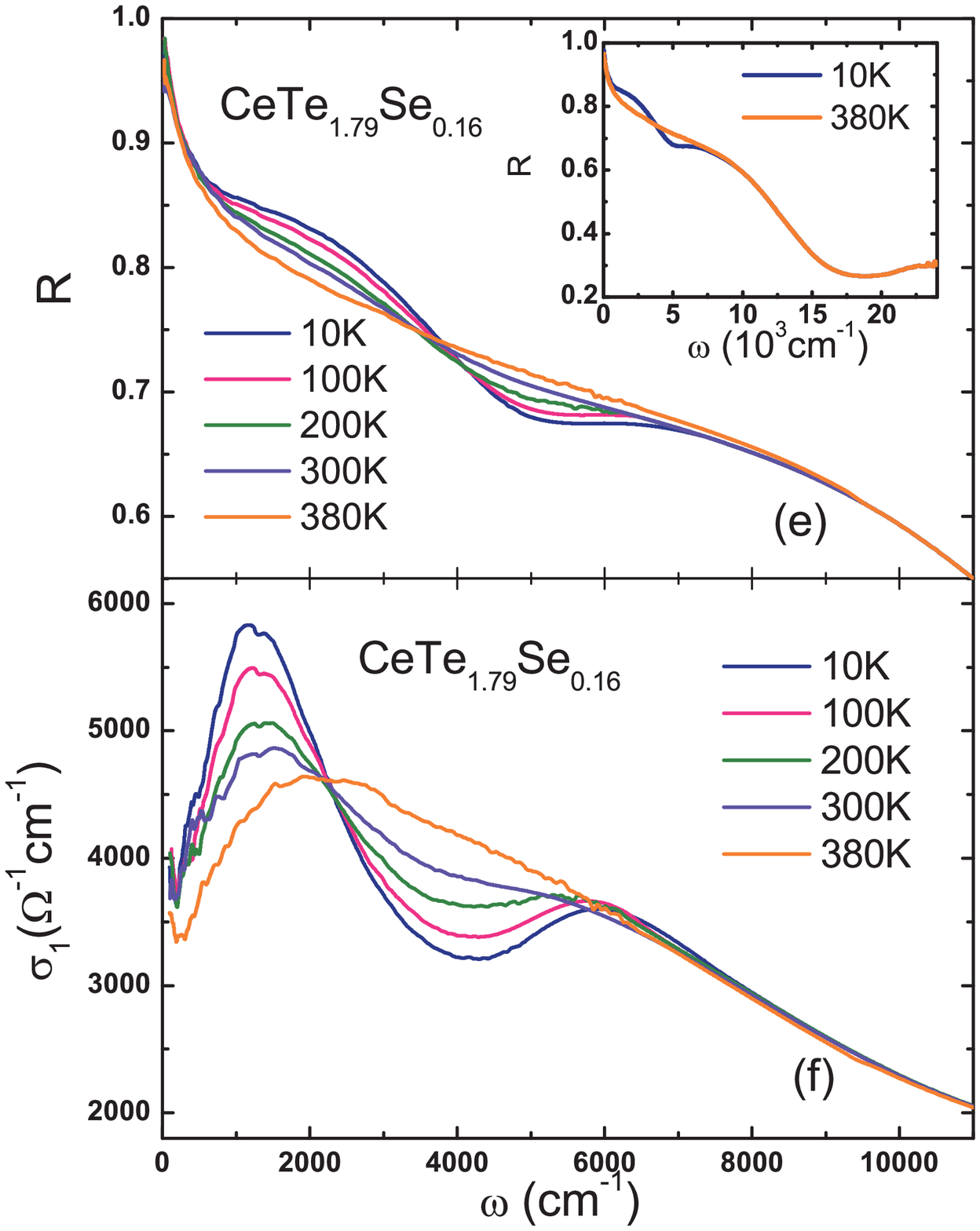}
\caption{(Color online)Left panel:(a) The temperature-dependence reflectivity of LaTe$_{1.95}$ below 15000 \cm.The inset shows the R($\omega$) at two representative temperatures up to 25000 \cm .(b) The frequency dependence  of the real part of the conductivity at different temperatures. Inset:$\sigma_1(\omega)$ at 10 K and 300 K over a broad frequency range.Middle and right panels are for CeTe$_{1.95}$ and CeTe$_{1.79}$Se$_{0.16}$ respectively.}
\end{figure*}

Figure 3 show the optical reflectance R($\omega$) and real part of conductivity
$\sigma_1(\omega)$ of LaTe$_{1.95}$, CeTe$_{1.95-x}$Se$_x$ (x=0 and 0.16), respectively. The insets of upper panels show
R($\omega$) in the expanded rang up to 25000 \cm. The
R($\omega$) of the those compounds have high values at low
frequency and decreases rapidly with increasing frequency, forming
plasma edges near 18000 \cm. The experimental results demonstrate that the
materials are metallic even in the deep CDW state, consistent with
the dc resistivity measurements.

Two significant features exist in the optical
spectra of those samples. The first one is the very strong dip structure
in R($\omega$) in the near infrared region $\sim$5000 \cm, which becomes more pronounced upon cooling.
This leads to a remarkable peak structure in the conductivity spectra $\sigma_1(\omega)$ at higher energy scale.
For the three samples, the conductivity peak is the most prominent for the LaTe$_{1.95}$
crystal. It locates near 8000 \cm at 300 K, and shifts to further higher energy scale $\sim$ 8500 \cm
as the temperature decreases to 10 K. For the pure and Se-doped CeTe$_{1.95}$ samples,
the peak feature appears at lower energy scales and also becomes less pronounced.
The spectra provide optical evidence for the presence of an
energy gap, which could be ascribed to the CDW order. Because of the
"type-I coherent factor" for density wave order which gives rise to a characteristic
peak structure just above the energy gap in optical conductivity, the peak position
in $\sigma_1(\omega)$ could be identified as the energy scale of CDW gap.\cite{HuSDW,electrondynamic}
The measurement results
are in agreement with earlier reports.\cite{Lavagnini,optCeTe2} In ARPES measurements, the inner FSs
centered at $\Gamma$ point were found to be almost fully gapped with E$_g$=600 meV
(roughly 4800 \cm) for LaTe$_{1.95}$ \cite{fisherRTe2}. Since the ARPES measurement probes the gap
relative to the Fermi level, while the optical measurement
detects the excitation from occupied to unoccupied states, the gap value by optics
should double the gap size probed by ARPES. Roughly, the gap values are consistent
with earlier ARPES experiments \cite{fisherRTe2}.
For the Se-doped CeTe$_{1.95-x}$Se$_x$ (x=0.16), since a resistivity
jump is observed near 345 K, the CDW order
is believed to be formed only below this temperature. Indeed,
we found an absence of this gap feature in reflectance
and conductivity spectra at measurement temperature 380 K.

The second very strong structure is the presence of another prominent
peak in conductivity spectra $\sigma_1(\omega)$ at much lower energy scale, between
800-1500 \cm for LaTe$_{1.95}$ and CeTe$_{1.95-x}$Se$_x$ (x=0, 0.16). This peak
structure was also observed in earlier optical measurements
and was assigned to an energy gap as well\cite{optCeTe2}. Indeed, ARPES measurements revealed
energy gap features roughly at the half of above values. Nevertheless, we noticed that
the peak positions in $\sigma_1(\omega)$ moves slightly towards
the lower energy as temperature decreases for those compounds. Such
temperature dependent shift is opposite to the expectation of CDW energy gap
formation. On this account, the low-energy peak structure could not be
ascribed to the formation of a small CDW gap. This conclusion is further strengthened by the
measurement on Se-doped CeTe$_{1.95-x}$Se$_x$ (x=0.16). For this sample,
the CDW phase transition is already suppressed to 345 K. In accordance with this suppression,
the energy gap feature near 6000 \cm is not visible in our optical
measurement at 380 K. However, the low-energy peak feature
is still present at all measurement temperatures. If the low-energy peak feature
is developed from CDW order, it should be more easily suppressed by the Se doping.
Considering the fact that the
Te deficiencies are always present in the RTe$_{2-x}$ samples and the conduction
electrons are from the 5p electrons of square Te layers, we expect that the disorder-driven
electron localization effect is the dominant contribution to the formation of the low-energy peak
structure in $\sigma_1(\omega)$. It remains to be investigated whether or not the
electron localization effect, which leads to the peak structure in $\sigma_1(\omega)$, could
results in a gap-like feature in ARPES measurements.
As we shall elaborate blow, the localization modified
Drude model could reasonably reproduce the spectral feature of the experimental data.

Besides the above two very strong spectral structures, we also noticed a weak feature
near 10000 \cm in R($\omega$) for both LaTe$_{1.95}$ and CeTe$_{1.95}$ compounds,
which is more clearly visible at low temperature and
leads to a shoulder in $\sigma_1(\omega)$ spectra at slightly higher frequencies.
Different from the notable shift of peak corresponding to CDW orders,
the shoulder positions show little change at varied temperatures. It is likely that this feature
comes from the inter-band transitions.

\section{\label{sec:level2}Analysis from the localization modified Drude model}

\begin{table*}[htbp]
\begin{center}
\newsavebox{\tablebox}
\begin{lrbox}{\tablebox}
\begin{tabular}{p{70pt}cccccccccccc}

\arrayrulewidth=1pt
Sample&$\omega_{p}$&$\gamma_{D}$&$k_{F}\lambda$&$\omega_1$&$\gamma_1$&$S_1$&$\omega_2$&$\gamma_2$&$S_2$&$\omega_3$&$\gamma_3$&$S_3$\\[4pt]
\hline
LaTe$_{1.95}$(10K)  &21&1.5&1.3&8.5&2&27&10&5&30&23&40&60\\[4pt]
LaTe$_{1.95}$(300K)  &21.3&2&1.07&8&2.5&24.7&10&5.3&29&23&40&60\\[4pt]
CeTe$_{1.95}$(10K)  &26&1.3&1.18&6.8&2.4&22&8.8&5.6&29&22&40&65\\[4pt]
CeTe$_{1.95}$(300K)  &27&2&1.12&6.2&3.4&21&8.3&5.7&28&22&40&65\\[4pt]
CeTe$_{1.79}$Se$_{0.16}$(10K)  &23&1.6&0.9&6&3.2&14&9&14.7&36&-&-&-\\[4pt]
CeTe$_{1.79}$Se$_{0.16}$(380K)  &34&4&1.14&-&-&-&9&19&32&-&-&-\\[4pt]
\hline \hline
\end{tabular}
\end{lrbox}
\caption{Temperature dependence of the plasma frequency $\omega_p$
,scattering rate $\gamma_D$=1/$\tau_D$ and order parameter $k_{F}\lambda$ of the LMD term, the
resonance frequency $\omega_i$, the width $\gamma_i$=1/$\tau_i$
and the square root of the oscillator strength $S_i$ of the
Lorentz component(all entries in 10$^3$ \cm). The Lorentz term in lowest energy
responsible for the CDW order and the others responsible for inter-band transition.}
\scalebox{1.0}{\usebox{\tablebox}}
\end{center}
\end{table*}

Let us now analyze the evolution of the itinerant carriers and
the CDW gap excitations in a quantitative way. As stated above, we try to use the localization
modified Drude (LMD) model, instead of a simple Drude term, to analyze the low frequency
conductivity spectra, since the former is more appropriate in a carriers-localization
system.\cite{GLi,Mott,handbook} The high frequency interband transitions and energy
gap excitations could be described by the Lorentz components. Within the LMD and Lorentz approach,
the dielectric function would consist of two parts:
\begin{equation}
\epsilon(\omega)=\epsilon_{LMD}(\omega)+\sum_{i=1}^{N}\frac{S_i^2}{\omega_i^2-\omega^2-i\omega/\tau_i}.
\label{chik}
\end{equation}
and
\begin{equation}
\epsilon_{LMD}(\omega)=\epsilon_\infty-{{\omega_p^2}\over{\omega^2+i\omega/\tau_D}}\left [1- \frac{C}{\left(k_{F}\lambda\right )^2 }\left ( \sqrt{\frac{3}{\omega \tau _{D}}} -\left ( \sqrt{6}-1 \right )\right ) \right ].
\label{chik}
\end{equation}
Here, the first term in expression (1) is the LMD component, the second term is the Lorents components.
$\epsilon_\infty$ is the dielectric constant at high energy, $\omega_{p}$ the plasma frequency, $k_{F}$ the Fermi wave vector, $\lambda$ the mean free path and $\emph{C}$ a universal constant($\sim1$). The model was found to reproduce
the conductivity spectra fairly well. As examples, we show in Fig. 4 the
experimental data together with the fitting curves
for CeTe$_{1.95-x}$Se$_x$ (x=0 and 0.16) samples at 10 K, respectively.
In Table 1 we list the fitting parameters for the three different
samples at 300 (or 380) and 10 K, respectively. The disorder parameter ($k_{F}\lambda$) is
in general greater than 1, which is in the metallic side of the
metal-insulator transition in terms of Ioffe-Regel criterion.
The model yields consistent values with the dc conductivity
at the zero frequency limit where it takes the form for the metallic conduction,
\begin{equation}
\sigma _{LMD}(0)=\frac{\omega_{p}^2}{4\pi \gamma}[1-\frac{1}{(k_{F}\lambda)^2 }].
\label{chik}
\end{equation}
This expression could account for the reduction of conductivity due to localization effect
when $k_{F}\lambda > 1 $.\cite{Homes} Apparently, the LMD model is more suitable
in describing the carrier response in the infrared region as it can account
for the disorder effect presented in the samples. Nevertheless, it
should be remarked that the ($k_{F}\lambda$)
parameter becomes slightly smaller than 1 for CeTe$_{1.95-x}$Se$_x$ (x=0.16) sample at low temperature,
indicating further enhanced localization effect. This effect could be naturally attributed
to the random substitutions of Te sites by Se in the conducting Te layers,
which drives the sample into the non-metallic side
of the Ioffe-Regel criterion. The result is consistent with the semiconducting dc resistivity behavior.
In this circumstance, the LMD model is no longer valid for this sample.

The above analysis indicates that the free carrier response can be described by the LMD component.
We found that $\omega_p$ of both LaTe$_{1.95}$ and CeTe$_{1.95}$ decreases very slightly from room
temperature to 10K. On the other hand, the scattering rate ($\gamma$ =
1/$\tau$) decreases even faster. However, for CeTe$_{1.95-x}$Se$_x$ (x=0.16), the plasma
frequency $\omega_p$ = 33600 \cm at 380 K reduces to 23000 \cm at 10 K.
The square of plasma frequency $\omega_p$ is proportional to the effective carrier
density \emph{n}/\emph{m}$^{\ast}$ (where \emph{m}$^{\ast}$ is the effective carrier mass).
This result could be interpreted as the formation of the partial CDW gap which removes
those electrons near $E_{F}$ that
experience stronger scattering, leading to a reduction of both conducting carrier density and
the scattering rate due to the reduction of scattering channels.

\begin{figure}[htbp]
\includegraphics[width=8cm,clip]{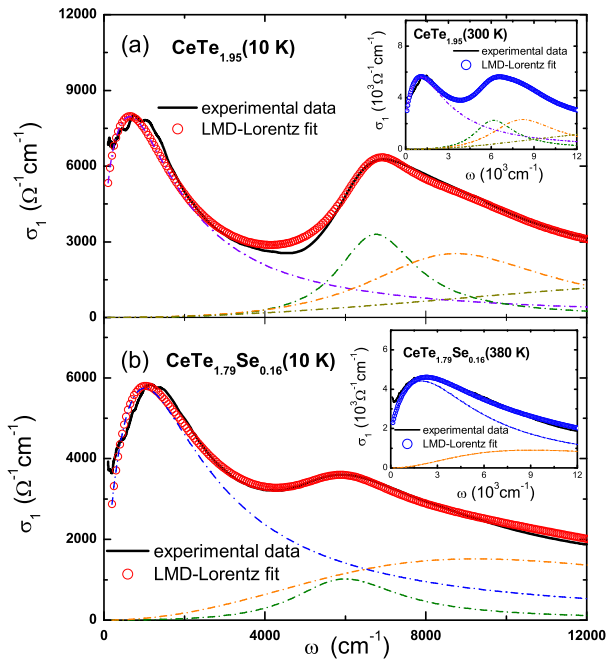}
\caption{(Color online) The experimental data of
$\sigma_1(\omega)$ at 10 K and the LMD-Lorentz fit results for (a) CeTe$_{1.95}$ and (b) CeTe$_{1.79}$Se$_{0.16}$.
The dashed curves displayed at the bottom in each panel are fitting components in the LMD-Lorentz analysis.
The one at the lowest energy scale is the LMD component, the others at the higher energy scales are Lorents components.
Inset shows the corresponding results at 300K or 380K.}
\end{figure}

Our study indicates that the Te-vacancies or disorder effect play the key role in the
low-energy charge excitations of rare-earth ditellurides RTe$_{2-x}$. This is different from the
extensively investigated rare-earth tri-telluride RTe$_3$ where Te vacancies were
usually not detected. We would also like to remark that the disorder effect could also dramatically
affect the CDW phase transitions.

As already discussed above, the CDW energy gap could be identified from the peak position of
the first Lorentz oscillation in the conductivity spectrum. At the lowest temperature, the Lorentz peak
locates near 8500 \cm for LaTe$_{1.95}$ and 6800 \cm for CeTe$_{1.95}$, respectively.
Those values are larger than the corresponding values of rare-earth tri-tellurides. Such
large energy gap values would imply that the compounds are already deeply in the CDW state even
at room temperature. With Se doping, CDW energy gap is reduced and
the CDW order is suppressed. However, the reduction of the energy gap is small.
In comparison with the undoped sample CeTe$_{1.95}$, the Lorentz peak, which
locates at 6000 \cm, is shifted by only 800 \cm for CeTe$_{1.79}$Se$_{0.16}$
sample. Even if we assume that the
ratio of the 2$\Delta/k_BT_c\approx8$, a number higher than BCS value but still often seen
in strongly coupling materials \cite{HuSDW}, we expect that the CDW transition temperature would be still
higher than 1000 K. In reality, the CDW transition temperature appears at 345 K.
To our knowledge, the CDW transition temperature close to room temperature in the RTe$_2$
system has never been observed before. Our study suggests that, in the heavily disordered
system, the ratio of the CDW energy gap over the transition temperature does not follow
the value as normally expected from the BCS mean-field theory for density wave instability.
Compared with the undoped samples, the
peak structure becomes much weaker. Our observations seem to indicate that the
disorder or localization effect arising from Te vacancies or Se substitutions affects the
the CDW transition temperature more radically than the energy gap.
We also emphasize that our results are strongly against
the conclusions drawn by Garcia \emph{et al.} based on ARPES study that CDW gap size
decreases dramatically as the number of the Te layers reduces from two (RTe$_3$) to one (RTe$_2$)
and the ReTe$_2$ would be examples of semiconducting quasi-2D CDW material due to
the gapping of the entire Fermi surfaces.

\section{\label{sec:level2}CONCLUSIONS}
To conclude, we have performed an optical study on the
single-crystals of LaTe$_{1.95}$, CeTe$_{1.95}$ and CeTe$_{1.79}$Se$_{0.16}$, belonging to the layered quasi-two-dimensional charge density wave systems.
Our measurement revealed clearly the formation of partial energy gaps
at rather high energy levels: 2$\Delta\sim$ 8500 \cm ($\sim$1.06 eV) for
LaTe$_{1.95}$, and 6800 \cm ($\sim$0.84 eV) for
CeTe$_{1.95}$. A small fraction of Se substitutions for Te dramatically weaken the CDW
order and suppress the phase transition temperature. As a result,
the CDW phase transition was observed, for the first time, close to room temperature in the
rare-earth ditelluride system. Our study also revealed that the low energy
excitations of the compounds are dominantly contributed
by the disorder effects due to the presence of Te vacancies in
conducting Te layers. The localization modified Drude model can
account for the low-frequency charge response fairly well. The spectral features are distinctly different
from a semiconducting CDW state with fully gapped Fermi surfaces.

\begin{center}
\small{\textbf{ACKNOWLEDGMENTS}}
\end{center}
This work was supported by the National Science Foundation of
China (10834013, 11074291), and the 973 project of the Ministry of
Science and Technology of China (2011CB921701, 2012CB821403).


\begin{references}


\bibitem{G.Gruner} G. Gr\"{u}ner, \emph{Density Waves in Solids} (Addison-Wesley, Reading, MA, 1994).

\bibitem{structure2} E. DiMasi, B. Foran, M. C. Aronson, and S. Lee, Chem. Mater. \textbf{6}, 1867 (1994).

\bibitem{ARPESCeTe3400meV} V. Brouet, W. L. Yang, X. J. Zhou, Z. Hussain, N. Ru, K.Y. Shin, I. R. Fisher, and Z. X. Shen, Phys. Rev. Lett. \textbf{93}, 126405 (2004).

\bibitem{material} N. Ru and I. R. Fisher, Phys. Rev. B \textbf{73}, 033101 (2006).

\bibitem{TEMRTe3} E. DiMasi, M. C. Aronson, J. F. Mansfield, B. Foran, and S. Lee,
Phys. Rev. B \textbf{52}, 14516 (1995).

\bibitem{Kikuchi} Kikichi, J. Phys. Soc. Jpn. \textbf{67}, 1308 (1998).

\bibitem{B.I.Minsuper} J. H. Shim, J.-S. Kang, and B. I. Min, Phys. Rev. Lett. \textbf{93}, 156406 (2004).

\bibitem{J.Lcaculate} J. Laverock, S. B. Dugdale, Zs. Major, M. A. Alam, N. Ru, I. R. Fisher, G. Santi, and E. Bruno,
Phys. Rev. B \textbf{71}, 085114 (2005).

\bibitem{fisherRTe2} K. Y. Shin, V. Brouet, N. Ru, Z. X. Shen and I. R. Fisher, Phys. Rev. B \textbf{72}, 085132 (2005).

\bibitem{DiMasi} E. DiMasi, B. Foran,M. C. Aronson, S. Lee, Phys. Rev. B \textbf{54}, 13587 (1996).

\bibitem{superconduct} M. H. Jung, A. Alsmadi, H. C. Kim, Yunkyu Bang, K. H. Ahn, K. Umeo, A. H. Lacerda, H. Nakotte,
H. C. Ri, and T. Takabatake, \textbf{67}, 212504 (2003).


\bibitem{optCeTe3} B. F. Hu, P. Zheng, R. H. Yuan, T. Dong, B. Cheng, Z. G. Chen, and N. L. Wang, Phys. Rev. B \textbf{83}, 155113 (2011).

\bibitem{optErTe3} B. F. Hu, B. Cheng, R. H. Yuan, T. Dong, A. F. Fang, W. T. Guo, Z. G. Chen, P. Zheng, Y. G. Shi, and N. L. Wang, Phys. Rev. B  \textbf{84}, 155132 (2011).

\bibitem{ARPESLaTe2} D. R. Garcia, G.-H. Gweon, S. Y. Zhou, J. Graf, C. M. Jozwiak, M. H. Jung, Y. S. Kwon, and A. Lanzara, Phys. Rev. Lett. \textbf{98}, 166403 (2007).

\bibitem{xrd} W. B. Pearson, Zeitschrift f\"{u}r Kristallographie\textbf{171}, 23 (1985).

\bibitem{Resist} Y. S. Kwon and B. H. Min, Physica B \textbf{281¨C282}, 120 (2000).

\bibitem{HuSDW} W. Z. Hu, J. Dong, G. Li, Z. Li, P. Zheng, G. F. Chen, J. L. Luo, and N. L. Wang, Phys. Rev. Lett. \textbf{101}, 257005 (2008).

\bibitem{electrondynamic} M. Dressel, and G. Gr\"{u}ner, \emph{Electrodynamics of Solids} (Cambridge, Reading, 2002).

\bibitem{Lavagnini} M. Lavagnini, A. Sacchetti, L. Degiorgi, K. Y. Shin, and I. R. Fisher,
Phys. Rev. B \textbf{75}, 205113 (2007).

\bibitem{optCeTe2} K. E. Lee, C. I. Lee, H. J. Oh, M. A. Jung, B. H. Min, H. J. Im, T. Iizuka, Y. S. Lee,
S. Kimura, and Y. S. Kwon, Phys. Rev. B  \textbf{78}, 134408 (2008).

\bibitem{GLi} G. Li, P. Zheng, N. L. Wang, Y. Z. Long, Z. J. Chen, J. C. Li and M. X. Wan, J. Phys.: Condens. Matter \textbf{16} 6195 (2004).

\bibitem{Mott} N. F. Mott, Metal-Insulator Transitions (Taylor \& Francis, London, 1990).

\bibitem{handbook} R. Menon, C. O. Yoon, D. Moses, and A. J. Heeger, 1998 Handbook of Conducting Polymers 2nd edi, vol 2 (New York: Dekker).

\bibitem{Homes} G. Tzamalis, N. A. Zaidi, C. C. Homes, and A. P. Monkman, Phys. Rev. B \textbf{66}, 085202 (2002).


\end{references}
\end{document}